# Design of Compandor Based on Approximate the First-Degree Spline Function

Lazar Velimirović[1], Zoran Perić[2], Miomir Stanković[3], Jelena Nikolić[2]
[1]Mathematical Institute of the Serbian Academy of Sciences and Arts, Kneza Mihaila 36,
11001 Belgrade, Serbia
[2]Faculty of Electronic Engineering Niš, Aleksandra Medvedeva 14, 18000 Niš, Serbia
[3]Faculty of Occupational Safety Niš, Čarnojevića 10 A, 18000 Niš, Serbia
velimirovic.lazar@gmail.com

*Abstract*—In this paper, the approximation of the optimal compressor function using spline function of the first-degree is done. For the companding quantizer designed on the basis of the approximative spline function of the first-degree, the support region is numerically optimized to provide the minimum of the total distortion for the last segment. It is shown that the companding quantizer with the optimized support region threshold provides the signal to quantization noise ratio that is very close to the one of the optimal companding quantizer having an equal number of levels.

## I. INTRODUCTION

In digital signal processing, quantization is the process of approximating a continuous range of values (or a very large set of possible discrete values) by a relatively-small set of discrete values. A device that performs quantization is called a quantizer [1]. Quantizers can be uniform and nonuniform. It is well known that uniform quantizers are suitable for signals that have approximately uniform probability density function (PDF) [1]. Since many of the real signals are characterized by continuous Laplacian random variable, which means that some input values are expected to be more often than the others, nonuniform quantization is more common case in practice. Nonuniform quantization can be realized through the process of companding, in which a specific compressor function is applied on an input signal [1]. Specifically, nonuniform quantization can be achieved by compressing the signal *x* using a nonuniform compressor characteristic $c(\cdot)$, by quantizing the compressed signal $c(x)$ employing a uniform quantizer, and by expanding the quantized version of the compressed signal using a nonuniform transfer characteristic $c^{-1}(\cdot)$ that is inverse to that of the compressor. The overall structure of a nonuniform quantizer consisting of a compressor, a uniform quantizer, and an expandor in cascade is called compandor or companding quantizer [1]. Although the optimal compressor function gives the maximum of the signal to quantization noise ratio (SQNR) at the reference variance at which the optimal companding quantizer is designed, the optimal companding quantizer is not widely used because it is very complicated to be realized practically [1]. In order to provide easier practical realization, approximation of the optimal compressor function is performed. Accordingly, in this paper, the approximation of the optimal compressor function using spline function of the first-degree, for Laplacian PDF is done.

A quantizer support region can be divided into a variety of ways. Unlike with the quantizers described in [2]-[5], the support region of the proposed quantizer model is not divided into segments of equal width. In [5], the approximation of the optimal compressor function using spline function of the first and second degree, for Laplacian PDF is done. As already mentioned, the support region of the quantizer described in [5] is divided into equal in width segments, each of which has an unequal number of cells. This results in a higher complexity of encoding and decoding procedure compared to the case where the number of cells per segments are equal. In order to overcome the mentioned problems with the realization of the optimal companding quantizer, and also to decrease the encoding and decoding complexity, in this paper we develop a new method of construction companding quantizers which introduces an equal number of cells within segments of unequal size. Also, in order to improve the performance of our companding quantizer, the method we propose introduces the optimization of the support region threshold. By designing the proposed quantizer based on the approximate spline function and with optimized support region threshold, SQNR that is close to that of the nonlinear optimal companding quantizer is obtained.

The rest of the paper is organized as follows: In section 2 the detailed description of spline functions of the first-degree and optimal compressor function is given. Design of the quantizer based on the approximate spline function of the first-degree is described in section 3. Also, the procedure for optimization of the optimal support region threshold is described in section 3. Finally, the achieved numerical results for the Laplacian source of unit variance are discussed in section 4.

Manuscript received April XX, 20XX; accepted April XX, 20XX.
This research was funded by a grant (No. XXX-00/0000) from the Research Council of Lithuania. This research was performed in cooperation with the Institution.

## II. APPROXIMATIONS USING SPLINE FUNCTIONS OF THE FIRST-DEGREE

In this paper, the approximation of the optimal compressor function using the spline function of the first-degree is done. The spline function is a function that consists of polynomial pieces joined together with certain smoothness conditions [6]. A simple example is the polygonal function (or spline of degree 1), whose pieces are linear polynomials joined together to achieve continuity [6]. In the theory of splines, the points $x_0, x_1,..., x_L$ at which the function changes its character are termed knots. Such a function appears somewhat complicated when defined in explicit terms. Accordingly, we consider the following definition of a linear polynomial $S(x)$ [6]:

$$S(x) = \begin{cases} S_0(x), & x \in [x_0, x_1] \\ S_1(x), & x \in [x_1, x_2] \\ \vdots \\ S_{L-1}(x), & x \in [x_{L-1}, x_L] \end{cases}, \quad (1)$$

where

$$S_i(x) = a_i x + b_i. \quad (2)$$

Obviously, $S(x)$ is a piecewise linear function. For the given knots $x_0, x_1,..., x_L$ and coefficients $a_0, b_0, a_1, b_1,..., a_{L-1}, b_{L-1}$, the evaluation of $S(x)$ at a specific $x$ performs by first determining the interval that contains $x$ and then by using the appropriate linear function for that interval. If the function $S$ defined by equation (1) is continuous, we call it a first-degree spline. It is characterized by the following three properties [6].

**Definition 1.** A function $S$ is called a spline of the first-degree if:
1. The domain of $S$ is an interval $[a, b]$.
2. $S$ is continuous on $[a, b]$.
3. There is a partitioning of the interval $a = x_0 < x_1 <_{...} < x_L = b$ such that $S$ is a linear polynomial on each subinterval $[x_i, x_{i+1}]$.

Outside the interval $[a, b]$, $S(x)$ is usually defined to be the same function on the left of $a$ as it is on the leftmost subinterval $[x_0, x_1]$ and the same on the right of $b$ as it is on the rightmost subinterval $[x_{L-1}, x_L]$ [6]. In other words, $S(x) = S_0(x)$ when $x < a$ and $S(x) = S_{L-1}(x)$ when $x > b$. Continuity of a function $f$ at a point $s$ can be defined by the condition [6]:

$$\lim_{x \to s^+} f(x) = \lim_{x \to s^-} f(x) = f(s). \quad (3)$$

Here, $\lim_{x \to s^+}$ means that the limit is taken over $x$ values that converge to $s$ from above $s$; that is, $(x-s)$ is positive for all $x$ values. Similarly, $\lim_{x \to s^-}$ means that the $x$ values converge to $s$ from below.

The first-degree spline, also called the polygonal function, is consisted of line segments that are connected so that given function is continuous. As alredy mentioned, the points where the function changes its shape are called knotes [6]. The approximate function $g(x)$, by which a nonlinear compressor function $c(x)$ is approximated in this paper, for the number of segments $L$, has the following form:

$$g(x) = \begin{cases} c(x_1) + m_1(x - x_1), & x \in [0, x_1] \\ c(x_i) + m_i(x - x_i), & x \in [x_{i-1}, x_i], \quad i = 2, \dots, L \end{cases}, \quad (4)$$

where $m_i$ is the coefficient of direction of the line given by the formula:

$$m_i = \frac{c(x_i) - c(x_{i-1})}{x_i - x_{i-1}}, \quad i = 1, \dots, L. \quad (5)$$

The optimal compressor function $c(x)$: $[-x_{\max}, x_{\max}] \to [-x_{\max}, x_{\max}]$ by which the maximum SQNR is achieved for the reference variance of an input signal having PDF $p(x)$ is defined as [1]:

$$c(x) = \begin{cases} x_{\max} \dfrac{\int_0^x p^{1/3}(x)dx}{\int_0^{x_{\max}} p^{1/3}(x)dx}, & 0 \leq x \leq x_{\max} \\[2ex] -x_{\max} \dfrac{\int_x^0 p^{1/3}(x)dx}{\int_{-x_{\max}}^0 p^{1/3}(x)dx}, & -x_{\max} \leq x \leq 0 \end{cases}. \quad (6)$$

Without diminishing the generality, in what follows the quantizer design will be done for the reference input variance of $\sigma_{ref}^2 = 1$.

## III. DESIGN OF COMPANDOR BASED ON APPROXIMATE THE FIRST-DEGREE SPLINE FUNCTION

This section provides us with a detailed description of the scalar compandor designed according the approximative spline function of the first degree. The support region threshold of the $N$-level companding quantizer is defined as follows [7]:

$$x_{\max} = \frac{3}{\sqrt{2}} \ln\left(\frac{N+1}{3}\right). \quad (7)$$

Let us assume, as in [2], that the total number of the reproduction levels per segments in the first quadrant is:

$$\sum_{i=1}^{L} \frac{N_i}{2} = \frac{N-2}{2}, \quad (8)$$

where the number of reproduction levels per segments, $N_i/2$, is determined from the following condition:

$$\frac{N_i}{2} = \frac{N}{2L}, \quad i = 1, \dots, L-1. \quad (9)$$

Then, the number of reproduction levels in the last segment, $N_L/2$, is:

$$\frac{N_L}{2} = \frac{N-2}{2} - (L-1)\frac{N}{2L}. \qquad (10)$$

The segment thresholds of our companding quantizer are determined as follows [1]:

$$x_i = c_i^{-1}\left(i\frac{N}{2L}\right), \quad i = 1,\ldots,L-1, \qquad (11)$$

where it obviously holds $x_L = x_{\max}$. Cells lengths per segments of the considered companding quantizer are given by:

$$\Delta_{i,j} = \frac{\Delta}{g_i'(x)}, \quad i = 1,\ldots,L, \ j = 1,\ldots,\frac{N_i}{2}, \qquad (12)$$

where

$$\Delta = \frac{2x_{\max}}{N-2}. \qquad (13)$$

Recall that $g(x)$ is the approximate function, by which we approximate the nonlinear optimal compressor function $c(x)$ (6). Denoted by $\Delta_{i,j}$ are the $j$-th cells lengths within the $i$-th segment. The cells thresholds of the considered quantizer are determined as follows:

$$x_{i,j}^{cell} = \frac{j\Delta}{g_i'(x)}, \quad i = 1, \ j = 1,\ldots,\frac{N_i}{2}, \qquad (14)$$

$$x_{i,j}^{cell} = \frac{c_i(x_i) + j\Delta}{g_i'(x)}, \quad i = 2,\ldots,L, \ j = 1,\ldots,\frac{N_i}{2}. \qquad (15)$$

The granular distortion for the proposed quantizer model is defined by [1]:

$$D_g = 2\sum_{i=1}^{L}\sum_{j=1}^{\frac{N_i}{2}} \frac{\Delta_{i,j}^2}{12} P_{i,j}, \quad i = 1,\ldots,L, \ j = 1,\ldots,\frac{N_i}{2}, \qquad (16)$$

where $P_{i,j}$ denotes the probability of the input sample $x$ of variance $\sigma^2$ belonging to the $j$-th cells in $i$-th segment. For the assumed Laplacian PDF of unit variance [1]:

$$p(x) = \frac{1}{\sqrt{2}\sigma} e^{-\frac{|x|\sqrt{2}}{\sigma}}, \qquad (17)$$

we can derive the following closed-form expressions for the probablities $P_{i,j}$:

$$P_{i,j} = \int_{x_{i,j-1}^{cell}}^{x_{i,j}^{cell}} p(x)dx = \frac{1}{2}\left[\exp(-\sqrt{2}x_{i,j-1}^{cell}) - \exp(-\sqrt{2}\,x_{i,j}^{cell})\right]. \qquad (18)$$

The overload distortion $D_o$ is defined by [1]:

$$D_o = 2\int_{x_{\max}}^{\infty}(x - y_N)^2 p(x)dx = \frac{1}{2}\exp(-\sqrt{2}x_{\max}) \qquad (19)$$

where $y_N$ is, as in [2] and [8], determined from the centroid condition:

$$y_N = \frac{\int_{x_{\max}}^{\infty} xp(x)dx}{\int_{x_{\max}}^{\infty} p(x)dx}. \qquad (20)$$

By determining the total distortion $D$ as a sum of the granular distortion $D_g$ (16) and the overload distortion $D_o$ (19), the signal to quantization noise ratio can be determined [1]:

$$\mathrm{SQNR} = 10\log\left(\frac{\sigma^2}{D_g + D_o}\right) = 10\log\left(\frac{\sigma^2}{D}\right). \qquad (21)$$

In order to improve the performance of the proposed companding quantizer, we propose a new method of construction the companding quantizer. Our method introduces the optimization of the support region threshold, i.e. the optimization of the approximate spline of degree 1 in the last segment. Numerical determination of the optimal support region threshold is performed respecting the criterion of minimum total distortion $D_L$ for the last segment:

$$D_L = 2\frac{\left(\frac{x_{\max}^{opt} - x_{L-1}}{N/L - 1}\right)^2}{12}\int_{x_{L-1}}^{x_{\max}^{opt}} p(x)dx + \frac{1}{2}\exp(-\sqrt{2}x_{\max}^{opt}). \qquad (22)$$

In other words, we perform optimization of $D_L$ with respect to the support region threshold. For the optimized support region threshold $x_{\max}^{opt}$, obtained in this way, we design approximate spline of degree 1 and we determine other parameters required to design the described companding quantizer. Numerical results that follows shows that by designing the proposed companding quantizer based on the approximate spline of degree 1 and with the optimized support region threshold, SQNR that is close to that of the nonlinear optimal companding quantizer is obtained.

### IV. NUMERICAL RESULTS AND CONCLUSIONS

Numerical results presented in this section are obtained for the case when the number of levels is equal to $N = 128$ and for the number of segments $2L = 8$ and $2L = 16$. In Fig. 1 and Fig. 2 the dependence of the distortion $D_L$ for the last segment of the proposed companding quantizer on the support region threshold $x_{\max}$ for the number of segments $2L = 8$ and $2L = 16$ is shown. Based on Fig. 1 it can be concluded that the minimum value of the total distortion $D_L$ of the proposed companding quantizer is achieved for the optimal value of the support region threshold $x_{\max}^{opt} = 6.78$.

Also, based on Fig. 2 it can be concluded that minimum value of the total distortion $D_L$ of the proposed companding quantizer is achieved for the optimal value of the support region threshold $x_{max}^{opt} = 7.28$. Table 1 shows the values of SQNR for the proposed companding quantizer and the case where the support region threshold is determined by (7), (SQNR$^F$), the values of SQNR for the proposed companding quantizer and the case where the support region threshold $x_{max}^{opt}$ is numerically optimized, (SQNR$^N$), and the values of SQNR of the optimal companding quantizer having $c(x)$ given by (6), (SQNR$^O$). The number of segments assumed is $2L = 8$ and $2L = 16$. Analyzing the results shown in Table 1 one can conclude that design of the proposed companding quantizer based on the approximate spline of degree 1, with the optimized support region $x_{max}^{opt}$, provides SQNR very close to that of the optimal companding quantizer having the same number of levels. In addition, given fixed number of levels $N = 128$, $2L = 8$ and $2L = 16$, numerical comparison reveals that the gain in SQNR achieved with the proposed quantizer model compared to the one proposed in [4] ranges up to 0.12 dB. Also, by comparing the performance of the proposed companding quantizer with the quantizer model having equidistant segment thresholds [3], it can be concluded that the proposed model is a very effective quantizer solution because it achieves a higher quality of the quantized signal (about 0.85 dB), for the same total number of levels $N = 128$ and for the number of segments $2L = 16$.

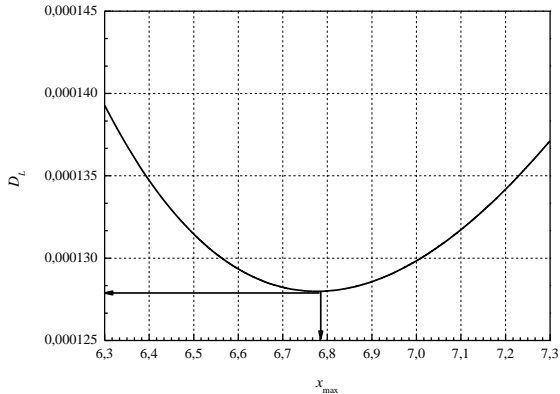

Fig. 1. Numerical determination of the optimal support region threshold for the number of segments $2L = 8$

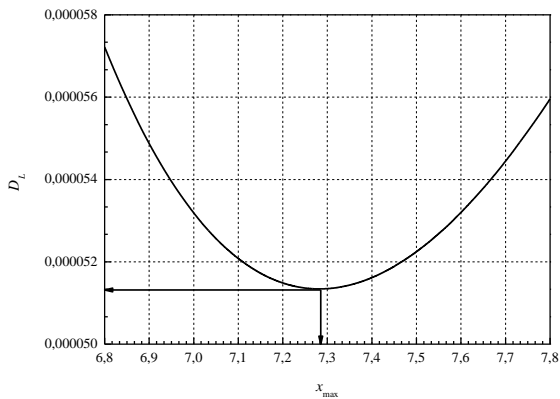

Fig. 2. Numerical determination of the optimal support region threshold for the number of segments $2L = 16$

TABLE I. THE VALUES OF SQNR$^F$, SQNR$^N$ AND SQNR$^O$ FOR THE NUMBER OF SEGMENTS $2L = 8$ AND $2L = 16$ AND $N = 128$.

| 2L | SQNR$^F$ [dB] | SQNR$^N$ [dB] | SQNR$^O$ [dB] |
|---|---|---|---|
| 8 | 34.19 | 34.76 | 35.71 |
| 16 | 35.21 | 35.41 | 35.71 |

ACKNOWLEDGMENT

This work is partially supported by Serbian Ministry of Education and Science through Mathematical Institute of Serbian Academy of Sciences and Arts (Project III44006) and by Serbian Ministry of Education and Science (Project TR32035).

REFERENCES

[1] N. S. Jayant and P. Noll, *Digital Coding of Waveforms: Principles and Applications to Speech and Video*. New Jersey: Prentice Hall, 1984, ch. 4-5.
[2] L. Velimirović, Z. Perić and J. NIkolić, "Design of novel piecewise uniform scalar quantizer for Gaussian memoryless source", *Radio Science*, vol. 47, no. 2, pp. 1-6, 2012.
[3] J. Nikolić, Z. Perić, D. Antić, A. Jovanović and D. Denić, "Low Complex Forward Adaptive Loss Compression Algorithm and ITS Aplication in Speech Coding", *Journal of Electrical Engineering*, vol. 62, pp. 19-24, 2011.
[4] J. Nikolić, Z. Perić, A. Jovanović and D. Antić, "Design of Forward Adaptive Piecewise Uniform Scalar Quantizer with Optimized Reproduction Level Distribution per Segments", *Elektronika IR Elektrotechnika*, vol 119, no. 3, pp. 19-22, 2012.
[5] L. Velimirović, Z. Perić, J. Nikolić and M. Stanković, "Design of Compandor Quantizer for Laplacian Source for Medium Bit Rate Using Spline Approximations", *Facta Universitatis*, vol. 25, no. 1, pp. 90-102, 2012.
[6] W. Cheney and D. Kincaid, *Numerical Mathematics and Computing, Sixth edition*. Belmont: Thomson Higher Education, 2008, ch 9.
[7] Z. Perić, M. Petković and M. Dinčić, "Simple Compression Algorithm for Memoryless Laplacian Source Based on the Optimal Companding Technique", *Informatica*, vol. 20, pp. 99-114, 2009.
[8] S. Na, "On the Support of Fixed-Rate Minimum Mean- Squared Error Scalar Quantizers for a Laplacian Source", *IEEE Transactions on Information Theory*, vol. 50, no. 5, pp. 937-944, 2004.